%% file: main.tex
\documentclass[twocolumn, english, aps,reprint, superscriptaddress,showpacs,longbibliography,showkeys, nofootinbib]{revtex4-2}
\usepackage{amsmath,amssymb,bbm,mathrsfs,bm,braket,color,graphicx,comment,xcolor,dsfont,multirow}
\usepackage[colorlinks,citecolor=blue,urlcolor=blue,linkcolor = blue]{hyperref}

\begin{document}

\date{\today}
\title{Fault-tolerant quantum computing with the parity code and biased-noise qubits}
\author{Anette Messinger}
\affiliation{Parity Quantum Computing GmbH, A-6020 Innsbruck, Austria}
\author{Valentin Torggler}
\affiliation{Parity Quantum Computing Germany GmbH, D-20095 Hamburg, Germany}
\author{Berend Klaver}
\affiliation{Parity Quantum Computing GmbH, A-6020 Innsbruck, Austria}
\affiliation{Institute for Theoretical Physics, University of Innsbruck, A-6020 Innsbruck, Austria}
\author{Michael Fellner}
\affiliation{Parity Quantum Computing GmbH, A-6020 Innsbruck, Austria}
\affiliation{Institute for Theoretical Physics, University of Innsbruck, A-6020 Innsbruck, Austria}
\author{Wolfgang Lechner}
\affiliation{Parity Quantum Computing GmbH, A-6020 Innsbruck, Austria}
\affiliation{Parity Quantum Computing Germany GmbH, D-20095 Hamburg, Germany}  
\affiliation{Institute for Theoretical Physics, University of Innsbruck, A-6020 Innsbruck, Austria}

\begin{abstract}
We present a fault-tolerant universal quantum computing architecture based on a code concatenation of biased-noise qubits and the parity architecture. The parity architecture can be understood as an LDPC code tailored specifically to obtain any desired logical connectivity from nearest-neighbor physical interactions. The code layout can be dynamically adjusted to algorithmic requirements on-the-fly. This allows for implementations with any desired code distance with a universal set of fault-tolerant gates. In addition to the previously explored tool-sets for concatenated cat codes, our approach features parallelizable interactions between arbitrary sets of qubits by directly addressing the parity qubits in the code. The proposed scheme enables codes with less physical qubit overhead compared to the repetition code with the same code distances, while requiring only weight-3 and weight-4 stabilizers and nearest-neighbor 2D square-lattice connectivity.  
\end{abstract}
\maketitle
\section{Introduction}

A long standing goal in quantum computation is to develop quantum computers which are able to solve real-world problems of practical interest which are classically intractable \cite{feynman2018simulating,shor1994algorithms, grover1996fast, farhi2014quantum, biamonte2017quantum, knill2005quantum}. A hurdle on the way to this goal is that quantum bits are inherently affected by noise, which limits the complexity of the quantum algorithms which can be performed without additional countermeasures. Thus, for implementing useful quantum algorithms the concepts of fault-tolerance in quantum computation \cite{Nielsen2011,shor1996fault, gottesman1997stabilizer, kitaev2003fault, breuckmann2021quantum} are indispensable. In fault-tolerant quantum computation, the quantum state is redundantly encoded in an error-correction code and fault-tolerant quantum operations are performed on the encoded quantum state. Fault-tolerant quantum computation comes with an overhead of resources like the number of qubits involved or the time needed to execute a quantum algorithm. Extensive research has been conducted on fault-tolerant quantum computation using the surface code ~\cite{horsman2012surface, litinski2019game, fowler2018low}. The surface code has a relatively high error threshold and can correct for phase-flip as well as bit-flip errors~\cite{fowler2009high}. As a downside, it requires a large qubit overhead, shifting implementations of complex problems to the far future.

Recently, there has been a strong focus on using qubits with biased noise for fault-tolerant quantum computation \cite{guillaud_repetition_2019,chamberland2022,ruiz2024}. This allows for protecting against one error type (for instance bit-flips) on the level of the physical implementation of the qubit while the second error type (for instance phase-flips) can be corrected by another error-correction code such as the repetition code~\cite{guillaud_repetition_2019, chamberland2022} or low density parity check (LDPC) codes with only one type of stabilizer \cite{ruiz2024}, or by a simpler biased quantum error-correction code, such as asymmetric surface codes~\cite{chamberland2022}. Since error-correction codes that are designed for (predominantly) correcting only a single error type have an improved encoding rate and simpler error-correction cycles, a biased-noise implementation saves resources, bringing fault-tolerant quantum computation into reach for near-term application.

In this paper, we propose to use the parity code \cite{Fellner2022universal} in combination with biased-noise qubits for fault-tolerant quantum computation. The parity code allows for error-correction of one type of error (e.g.\ bit flips) but at the same time, the introduced redundancy is used to enable easy and physically local implementations of logical many-body operations. The underlying stabilizer operators can be measured and manipulated on a platform with nearest-neighbor interaction on a 2D grid,  while still leading to a high encoding rate as compared to the repetition code. 

We show how to implement fault-tolerant logical gates within the parity code on a two-dimensional chip layout using the concepts of code deformation~\cite{messinger2023}. In particular, an efficient and parallelizable implementation of the controlled-$Z$-gate is introduced, alongside a set of universal logical gates. The parity code thus
offers an alternative path towards logical long-range connectivity and parallelization which is directly obtained from the choice of the stabilizer instead of the use of transversal gates and lattice surgery.

This paper is organized as follows: Section~\ref{sec:paritycode} introduces different forms of the parity code. Section~\ref{sec:concat-paritycode} introduces the concatenation of the parity code with the cat code. In Section~\ref{sec:fault-tolerant-gates} we propose fault-tolerant implementation of logical gates for the concatenation of the parity code with biased-noise qubits. Finally, we conclude our findings and give an outlook to possible follow-up research in Section~\ref{sec:conclusion}.

\section{The Parity Code}\label{sec:paritycode}
The parity code, which was originally proposed as a mapping for optimization problems~\cite{Lechner2015}, can be described as a stabilizer code~\cite{messinger2023, Fellner2022universal} and used for error correction of a single type of error.
The stabilizers are chosen such that in the code space, every physical qubit corresponds to either a single logical qubit or the parity of multiple logical qubits (i.e., their relative orientation along the $Z$-axis).

We index logical qubits and operators by single numbers, while physical qubits and operators are labelled by sets of such numbers. Logical operators are also denoted using a tilde, while operators without tilde refer to operators on physical qubits.
For the logical operators $\tilde Z$, the parity mapping can then be written in the form
\begin{equation}\label{eq:parityrelation}
    Z_{\{a,b,\dots\}}\ket{\psi} = \tilde Z_a \tilde Z_b \dots \ket{\psi},
\end{equation}
or more generally,
\begin{equation}
     Z_{\mathcal{L}}\ket{\psi} = \prod_{i \in \mathcal{L}} \tilde Z_i \ket{\psi},
\end{equation}
where  $\mathcal{L}$ is the set of the indices of the logical qubits whose parity information the correspondingly labelled physical qubit contains. Note that this mapping is opposite to what usually occurs in quantum error correction codes, where logical single-body operators are mapped to physical multi-body operators.
We call physical qubits whose label contains at least two logical qubits \textit{parity qubits}, and physical qubits corresponding to only a single logical qubit \textit{base qubits}\footnote{In previous works on the parity code these were also called `data qubits'. In order to avoid confusion with the definition of data qubits in error correction literature, we use the term `base qubit' in this work.}.

Such a mapping allows us to easily implement logical many-body entangling gates, for example GZZ gates~\cite{Bassler2022} 
\begin{equation}
    e^{\frac{i}{2} \sum_{i,j}A_{ij} \tilde Z_i \tilde Z_j}.
\end{equation}
If the code contains one parity qubit for each term in such an interaction, the whole operation can be implemented in a single time step by addressing each parity qubit individually with the corresponding single-qubit rotation.

For a given set of desired base qubits and parity qubits, a suitable stabilizer group can always be generated from operators 
\begin{equation}\label{eq:stabilizer_operator}
C = \prod_{\mathcal{L} \in S_C} Z_{\mathcal{L}}.
\end{equation}
Here, $S_C$ is a set of different physical qubit labels $\mathcal{L}$ such that when combining the logical qubit indices of all labels in the product, i.e., the symmetric difference of all ${\mathcal{L}\in S_C}$, we obtain the empty set,
\begin{equation}
    \triangle S_C = \emptyset,
\end{equation}
where $\triangle$ denotes the symmetric difference.
If a code state $\ket{\psi}$ satisfies our desired mapping, eq. \ref{eq:parityrelation}, it must also be a +1 eigenstate of all such stabilizer operators,
\begin{equation}
    C\ket{\psi} = \prod_{\mathcal{L} \in S_C} Z_{\mathcal{L}}\ket{\psi} = \prod_{\mathcal{L} \in S_C}\prod_{i \in \mathcal{L}} \tilde Z_i\ket{\psi}=\ket{\psi},
\end{equation}
as all logical $\tilde Z$ operators appear an even amount of times and therefore cancel out.

\begin{figure}
    \centering
    \includegraphics[width=.9\columnwidth]{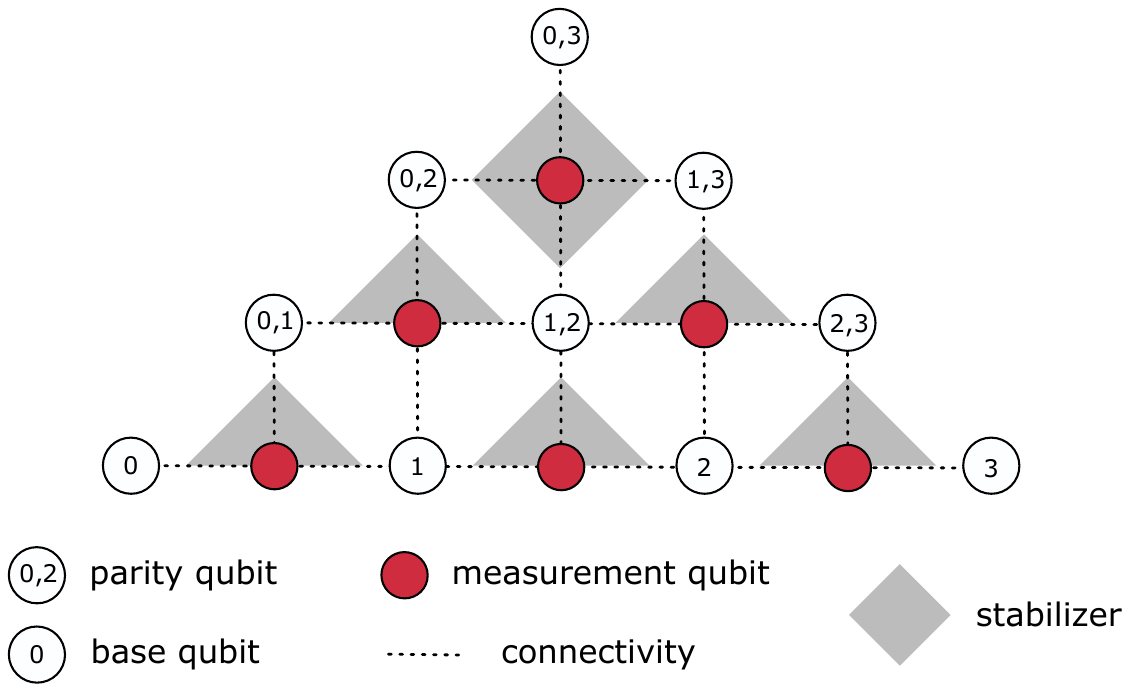}
    \caption{Parity Code in the LHZ Layout. All stabilizers are in the $Z$ basis and have weight 3 or 4. Physical $Z$ operations on base qubits (single-index label) translate directly to logical $Z$ operations on the corresponding logical qubits. Physical $Z$ operations on parity qubits (multi-index label) map to logical multi-body operations.}
    \label{fig:LHZlayout}
\end{figure}

As an example, let us consider the stabilizer operator containing the physical qubits with the labels in
\begin{equation}
S_C=\{\{0, 2\}, \{0, 3\}, \{1, 2\}, \{1, 3\}\}
\end{equation}
(see top-most square shape in Fig.~\ref{fig:LHZlayout}).
We can easily check that
\begin{equation}
    {\{0, 2\}} \triangle {\{0, 3\}} \triangle {\{1, 2\}} \triangle {\{1, 3\}} = \emptyset
\end{equation}
since each logical qubit index appears exactly twice.
The corresponding generator operator according to Eq.~\eqref{eq:stabilizer_operator} is 
\begin{equation}
    C=Z_{\{0, 2\}}Z_{\{0, 3\}}Z_{\{1, 2\}}Z_{\{1, 3\}}.
\end{equation}
We can easily check that  $C\ket\psi = \ket\psi$ is equivalent to the statement that a physical operator $Z_{\{0,3\}}$ has the same effect on a code state as the product of the other three operators,  

\begin{equation}
   Z_{\{0, 3\}}\ket\psi =  Z_{\{0, 2\}}Z_{\{1, 2\}}Z_{\{1, 3\}}\ket\psi
\end{equation}
and thereby define the logical operator on parity qubit $\{0,3\}$ indirectly via the parity of the others, $   \tilde Z_0 \tilde Z_2 \tilde Z_1 \tilde Z_2 \tilde Z_1 \tilde Z_3= \tilde Z_0 \tilde Z_3$.

It has been shown \cite{terhoeven2023} that, as long as additional (ancillary) base/parity qubits may be added, one can always find an arrangement of local stabilizers up to weight four for any set of desired qubit labels. A well-studied layout implementing logical all-to-all connectivity (i.e., containing a parity qubit for each possible pair-interaction) on a two-dimensional square lattice is the LHZ layout~\cite{Lechner2015} as shown in Fig.~\ref{fig:LHZlayout}. In this layout, $k$ base qubits are accompanied by ${\binom{k}{2} = k(k-1)/2}$ parity qubits and the same number of stabilizer generators.

\subsection{Logical operations}
In addition to the logical many-body operations which follow directly from the qubit labels as introduced in Eq.~\eqref{eq:parityrelation}, we can always define the single-body logical Pauli-Z operator accordingly as the product of operators on a set of physical qubits whose parity combines to the single desired logical qubit index,
\begin{equation}
    \tilde Z_i = \prod_{\mathcal{L} \in S_i} Z_{\mathcal{L}},
\end{equation}
where $S_i$ fulfills $\triangle S_i=\{i\}$. Such a set of labels can be found for any valid code which encodes logical qubit $i$. If the corresponding base qubit is part of the code, the product can be simplified to
\begin{equation}
    \tilde Z_i = Z_{\{i\}}.
\end{equation}

The logical Pauli-$X$ operator can be obtained from the product of physical Pauli-$X$ operators on all qubits whose label contain the logical qubit index,
\begin{equation}\label{eq:logicalX}
    \tilde X_i = \prod_{\substack{\mathcal{L}\\ i\in \mathcal{L}}} X_{\mathcal{L}}.
\end{equation}
In other words, we flip all qubits which in any form contain information about the logical qubit, either as base qubit or as parity qubit.

\subsection{Fault Tolerance}
As all stabilizers are Pauli-$Z$ products, the parity code alone corrects only bit-flip errors. In the LHZ layout, $k$ logical qubits are encoded in ${n=k(k+1)/2}$ physical qubits using ${n-k}$ stabilizers. 
Due to the regular structure of the LHZ layout, it is easy to check that all logical single-body $X$-operators have weight $k$, but also that any combination of them has a weight of at least $k$. This is due to the fact that any pair of logical $X$ operators overlaps on only one physical qubit, and that this qubit is unique to the pair. The product of $t$  logical $X$ operators therefore always has weight ${w(t) = t * k - t(t-1)}$, as each of the $t$ operators involves $k$ qubits, but there will be $\frac{t(t-1)}{2}$ qubits on which two operators overlap. In the range ${[ 1, k]}$, this function has its minima of ${w_{\rm min} = k}$ right at the range limits ${t=1}$ and ${t=k}$.
All code words are therefore separated by a Hamming distance of at least $k$, implying a code distance of ${d=k}$ (note that the distance can still be controlled independently of the total logical qubit number by creating multiple smaller LHZ layouts or truncating an LHZ layout). In the limit of large systems, the encoding rate $k/n$  as a function of code distance of the LHZ layout is twice that of the repetition code,
\begin{equation}
    \frac{k}{n}=\frac{d}{d(d+1)/2}\approx\frac{2}{d},
\end{equation}
i.e., only half as many physical qubits are required for the same code distance and logical qubit number.

\begin{figure*}
\centering
    \includegraphics[width=0.8\textwidth]{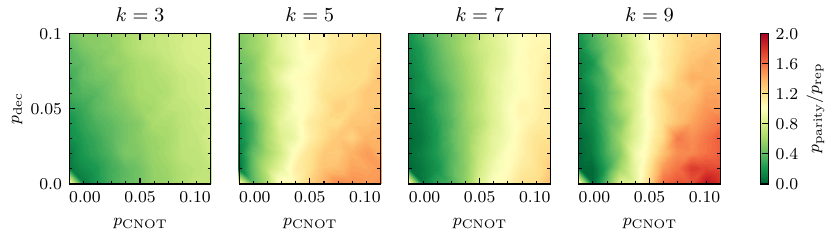}
    \caption{Ratio of the logical error rate $p_{\text{parity}}$ of the parity code using LHZ layout over the logical error rate $p_{\text{rep}}$ of $k$ repetition codes per stabilizer measurement round with equal number of physical qubits as a function of decoherence error probability $p_{\text{dec}}$ and CNOT error probability $p_{\text{CNOT}}$ for multiple logical system sizes $k$. The value at ${p_{\text{dec}} =p_{\text{CNOT}}=0}$ is set to $1$. The decoherence error represents bit-flip errors induced by interaction with the environment, while any phase-flip errors are neglected in this calculation. The parity code is advantageous to the repetition code in regimes with non-vanishing error rates and CNOT errors below approximately $5\%$. Data available at \cite{data}}
\label{fig:error_rates}
\end{figure*}

To compare the error-correction capabilities of this parity code and the repetition code we consider the same number of physical qubits and logical qubits for both codes. In particular, we compare the total logical error rate (the probability of at least one logical qubit being erroneous)  $p_{\text{parity}}$ of the $[n=k(k+1)/2, k, d=k]$ parity code in the LHZ layout to the total logical error rate $p_\text{rep}$ of $k$ repetition codes $[n=k(k+1)/2, k,  d=(k+1)/2]$, as a function of the number of logical qubits $k$.
Furthermore, we distinguish between error probabilities induced by CNOT gates $p_\text{CNOT}$ and bit-flip errors induced by interaction with the environment $p_\text{dec}$ during a stabilizer measurement round. Both codes are corrected using the Belief Propagation algorithm requiring only polynomial time with respect to the number of stabilizers~\cite{Pastawski_2016}. For various physical error probabilities,  Fig.~\ref{fig:error_rates} shows the ratio of the two logical error rates  $p_{\text{parity}}/p_{\text{rep}}$ for $10000$ trials in the case of perfect measurements. Note that, as we consider consecutive odd values for $k$, the corresponding repetition code can also have even distances ${(k+1)/2}$  in which case it performs worse in a comparison.

For the simulation, each parity or base qubit $i$ is assigned a total bit-flip error probability $p_{i}$, which is used to simulate an error by means of a Bernoulli trial. The total error probability is the probability of being affected by either error source, which we calculate as
\begin{equation}p_{i}=(1-p_\text{dec})g(p_\text{CNOT},w_{i})+p_\text{dec}(1-g(p_\text{CNOT},w_{i})), 
\end{equation}
where $g$ is a generating function dependent on the number $w_{i}$ of CNOT gates in which qubit $i$ is involved during a stabilizer measurement. All qubits are either involved in one, two or four CNOT gates, and the respective generating functions are 

\begin{equation*}
\begin{array}{ll}
g(p_\text{CNOT}, 1)&=p_{\text{CNOT}}\\
g(p_\text{CNOT}, 2)&=2 p_{\text{CNOT}}\left(1-p_{\text{CNOT}}\right)\\
g(p_\text{CNOT}, 4)&=4p_{\text{CNOT}}^3\left(1-p_{\text{CNOT}}\right)\\
&\,\,+4 p_{\text{CNOT}}\left(1-p_{\text{CNOT}}\right)^3\\
\end{array}
\end{equation*}

For low CNOT error rates the parity code is advantageous, here the result is determined by the fact that the LHZ layout has a larger code distance than the compared repetition code.
Only above a certain CNOT error rate the increased weight of the stabilizer operators and thus increased number of required CNOT gates in the LHZ layout make the repetition code beneficial.

While classical codes with even higher encoding rate (for a certain code distance) have been proposed~\cite{ruiz2024}, the LHZ layout requires only local stabilizers (of weight 3 or 4) which can be implemented on a 2D square lattice with nearest-neighbor connectivity. This type of stabilizer thus facilitates an easier technical implementation in a quantum device.

\subsection{Deformation of the Parity Code}

\begin{figure}
    \centering
    \includegraphics[width=\linewidth]{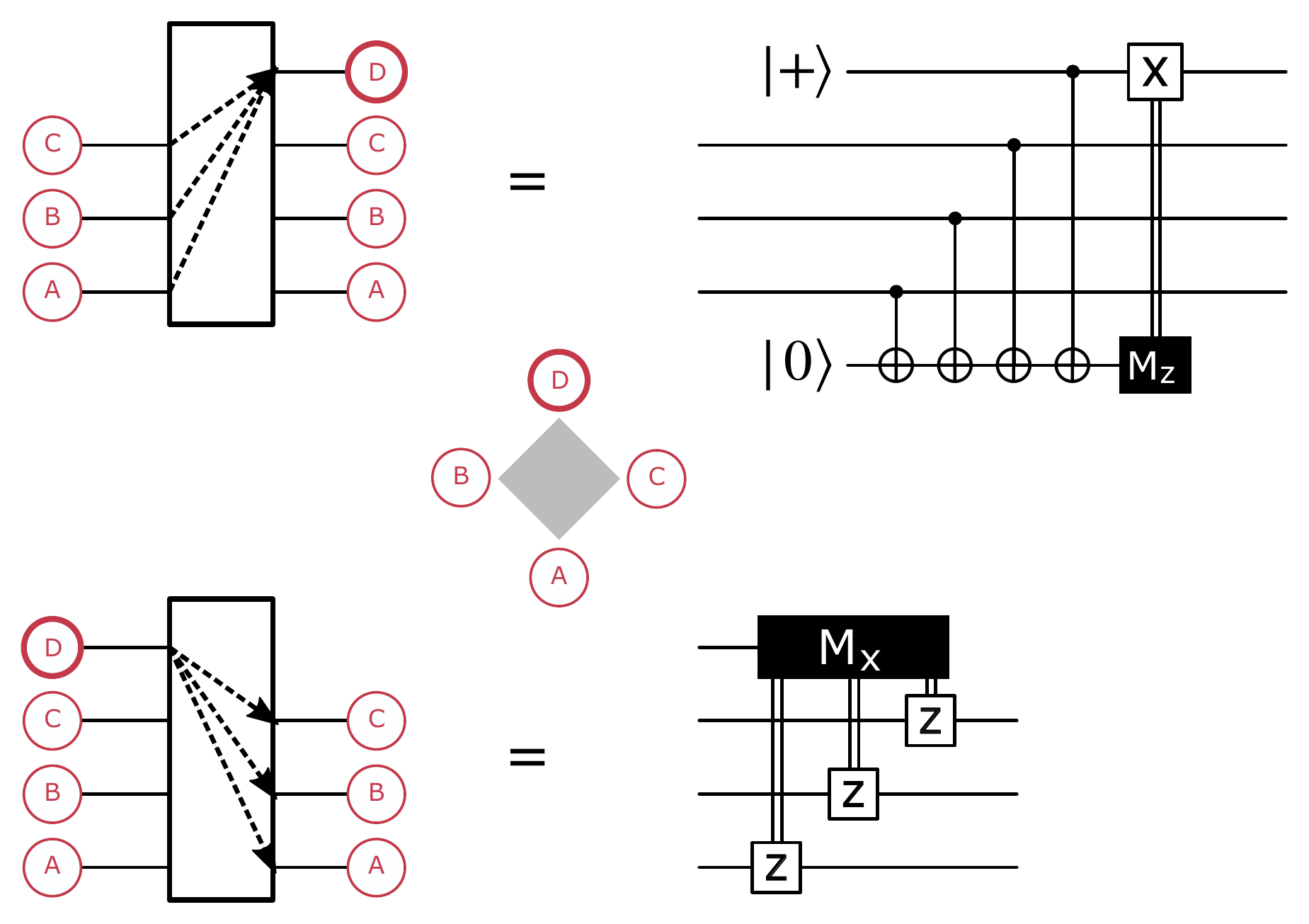}
    \caption{Protocols for adding (top) or removing (bottom) a parity qubit D which is in a stabilizer operator with qubits A,B and C (i.e., $D = A\triangle B\triangle C)$. Multiple addition or removal operations can be parallelized by
    pushing all corrections to the end and classically tracking any non-commuting effects of this (taking into account how they affect measurement outcomes, see \cite{messinger2023} for more details).  Both the protocols for addition and removal of qubits can be performed during syndrome measurement rounds. Note that, when a qubit is removed, the corresponding stabilizer is already excluded from that syndrome measurement round.}
    \label{fig:deformation}
\end{figure}

An important ingredient in constructing a fault-tolerant universal gate set is the deformation of the parity code, i.e., adding or removing base or parity qubits by changing the stabilizers, which can be done using CNOT gates~\cite{Fellner2022universal} or measurements and classical processing~\cite{messinger2023} as shown in Fig. \ref{fig:deformation}. 

For example, we can add a new qubit with label $\{1,3\}$ to the code by using already existing code qubits with labels $\{1,2\}$ and $\{2,3\}$ (or any other set of qubits whose labels combine to $\{1,3\}$ via symmetric difference). The new parity qubit can be added by initializing it in the state $|0\rangle$ and applying two CNOT gates controlled by the existing qubits $\{1,2\}$ and $\{2,3\}$, targeting the new qubit, thus imprinting the parity information of the control qubits to the new qubit. With the addition of a new qubit to the code must also come the addition of a new stabilizer operator linking the new qubit parity to the existing qubits. In this case this will be the operator $Z_{\{1,2\}}Z_{\{2,3\}}Z_{\{1,3\}}$.

The same process can also be achieved in a more parallelizable way by initializing the new qubit in the state $|+\rangle$, performing a stabilizer measurement of the operator\footnote{For easier readability and identification, we here already use the notation $Z_{\{1,3\}}$ to refer to the new qubit which we wish to become parity qubit $\{1,3\}$, even before we have accomplished this.} $Z_{\{1,2\}}Z_{\{2,3\}}Z_{\{1,3\}}$ and an $X$-flip on the new qubit if the measurement returned $-1$. With this construction, multiple new qubits can be added in a single round, even if some of the new qubits are needed to complete a local stabilizer for other new qubits. In such a case, all corrective $X$-flips can still be commuted to after the measurements such that only a single round of measurements and a single round of final $X$-flips is required. The required adjustments due to the effect which some of the $X$-flips would have had on following measurements can be calculated classically. Note that for a fully fault-tolerant implementation of this deformation in the presence of noise, multiple repetitions of the involved stabilizer measurement circuits might be necessary to determine the correct code space and identify faulty measurements with sufficient confidence. Besides, the use of additional flag qubits could be considered to further reduce the accumulation of high-weight errors.

Qubit $\{1,3\}$ can be removed from the code again by the reverse CNOT sequence it was added with. For a parallelizable approach, in this case the CNOT gates can be replaced by an $X$-measurement of qubit $\{1,3\}$ followed by conditional $Z$-flips on qubits $\{1,2\}$ and $\{2,3\}$ (or any other set of qubits which formed a stabilizer operator together with qubit $\{1,3\}$).
Except for very small code changes, it is beneficial to use the addition and removal protocols based on measurement and classical controls for all code changes, as they can be fully parallelized and combined with the usual stabilizer measurement cycles.

With these protocols, any desired set of parity/base qubits can be added to or removed from the code, as long as each addition or removal of a physical qubit in the code is accompanied by adding or removing a stabilizer generator acting on the affected qubit (and others). The only exception to this rule is the case where the respective qubit does not relate to other qubits of the code via their parity at all. In this case, a logical degree of freedom would be added or removed and thus no change of the stabilizer needed.

To allow for an easier implementation, we further restrict the protocol such that there is always a choice of stabilizer generator which can be all measured locally using nearest-neighbor operations. This means, each added qubit must relate to some of its neighboring qubits via their parity (share a stabilizer operator with them).

Following these rules, one can for instance add or remove a qubit with a label that represents the symmetric difference of the labels of other (ideally neighboring) qubits in the code, i.e., a ``new'' parity qubit. However, one can also add or remove ``copies'' of parity qubits, i.e., qubits with exactly the same label as others in the code. Also base qubits can be copied in that way, leading to a setup similar to a repetition code\footnote{In the extreme case of a setup consisting only of copies of base qubits, the stabilizer group is in fact exactly that of a repetition code.}. One can then make use of the similarities to repetition codes to apply strategies from concatenations of the cat code with repetition codes~\cite{guillaud2019,chamberland2022,guillaud2021} and, for example, implement transversal gates or gate teleportation schemes. Besides the vast toolbox of operations this gives us access to, code deformation can also be helpful to maintain a certain code distance. If for example a quantum algorithm or operation requires a certain (more limited) configuration of parity qubits on some part of the chip, any setbacks in terms of code distance (e.g., because fewer parity qubits are used in total) can be made up for by adding appropriate additional qubits on another part of the code. In the parity code, logical qubits (their $X$ operators) are de-localized across large areas of the physical device. This is useful not only to be able to implement interactions with other logical qubits locally on the appropriate parity qubits, but also allows more options for increasing the code distance: The weight of a logical $X$ operation can in principle be increased by adding new physical qubits (containing the logical index in their label) in the vicinity of any other physical qubits in that logical $X$ operation.

Note that, whenever the removal of a set of qubits would lead to a decrease of the code distance, other qubits to compensate this must be added before the removal so the code is never in a less-protected state. Furthermore, after adding new qubits to the code, multiple rounds of syndrome measurements might be necessary until their states can be trusted with enough confidence (in case of imperfect measurements). 

Different quantum operations might also require different code layouts. Whenever many interactions are required in parallel, for example in the form of the GZZ gate, a large number of parity qubits proves useful. Other types of interactions, in particular non-diagonal operations, are more difficult to implement in the presence of too many parity qubits. Here, it may be advantageous to encode a certain logical qubit simply in a number of copies of base qubits like in the repetition code and thus separate it from the other logical qubits. Both these layouts can be combined, such that some logical qubits can be in highly connected states while others are separated and easily accessible. In the following we show some exemplary layouts for different purposes to highlight the many options to use parity codes.

\begin{figure}
\begin{centering}
\includegraphics[scale=0.24]{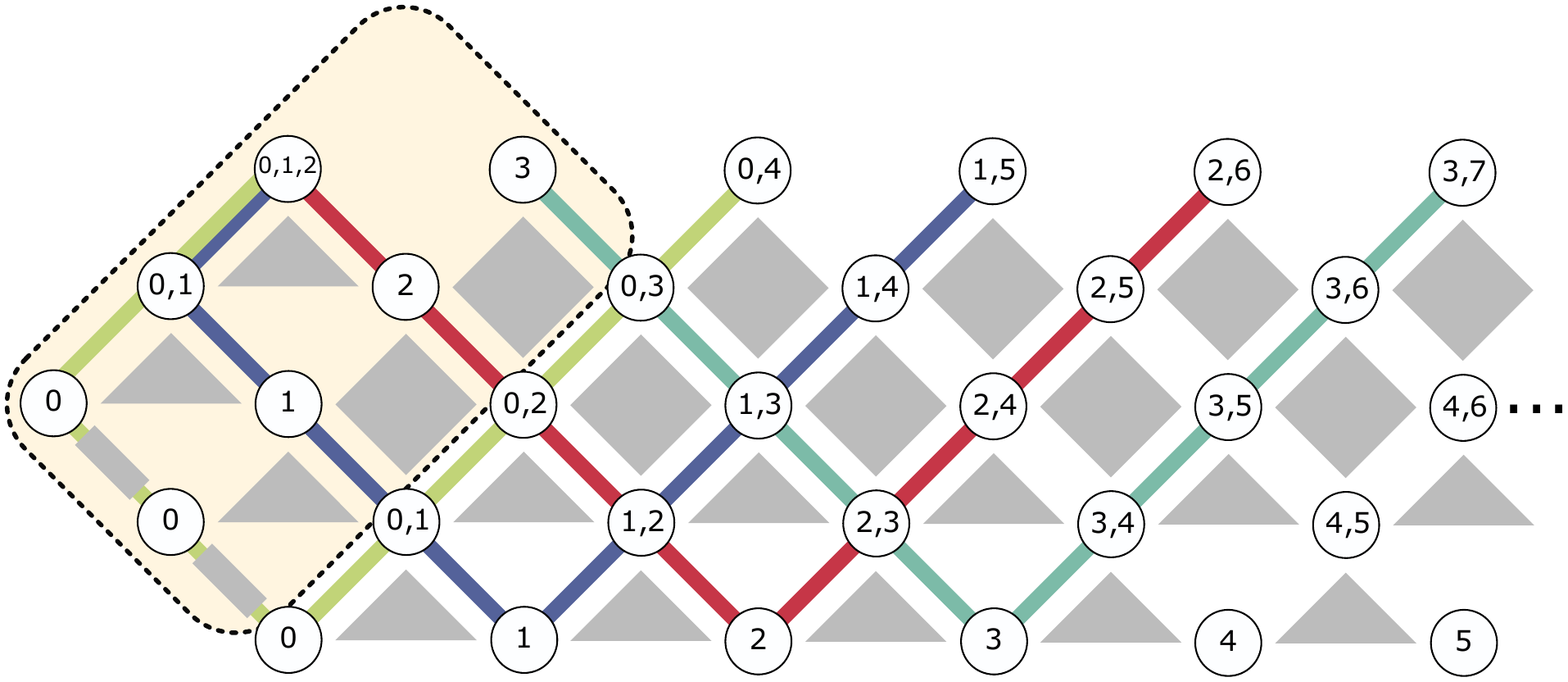}
\par\end{centering}
\caption{Layout of base qubits and parity qubits allowing fixed-range connectivity and constant
code distance $d=9$. The layout results from the full LHZ triangle~\cite{Fellner2022universal} with the tip (containing long-range parity
qubits) cut off. On the sides, an additional structure (highlighted
region) can be added to grow logical $X$ operators on the boundary to the same length
as the others and thereby retain higher code distance.\label{fig:limited-range}}

\end{figure}

If full connectivity and a code distance scaling with $k$ is not required at a certain stage of the algorithm, the system can for example be encoded in a setup with limited-range connectivity as shown in Fig.~\ref{fig:limited-range}.

\begin{figure}
\begin{centering}
\includegraphics[scale=0.24]{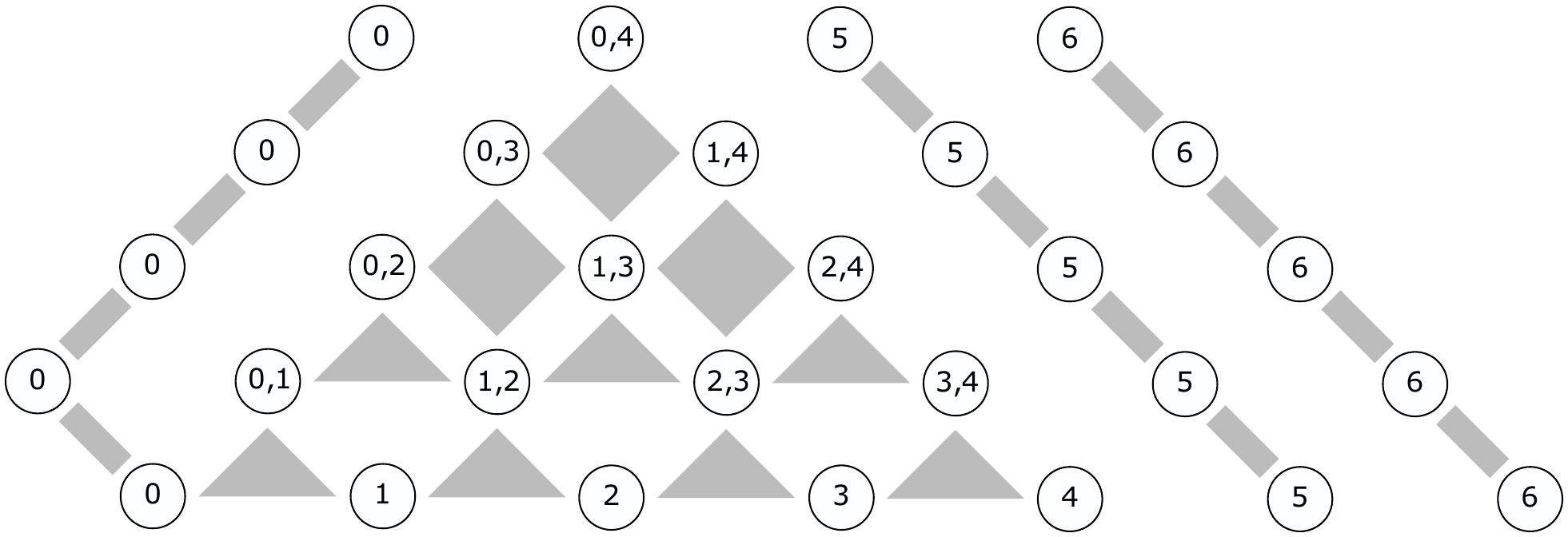}
\par\end{centering}
\caption{Example layout with code distance ${d=5}$, exhibiting high connectivity between logical qubits 0 to 4, while logical qubits 5 and 6 are decoupled from the others and encoded as in a repetition code. Logical qubit 0 has multiple base qubit copies in addition to parity qubits. This allows for using certain methods from repetition encodings, e.g., for implementing a fault-tolerant gate, while keeping the parity-mediated connectivity with other logical qubits.
\label{fig:mixed_layout}}

\end{figure}

In general, any layouts can be combined as long as a valid set of stabilizers is maintained. For example, a part of the code can be in a highly (or partially) connected layout, while other logical qubits are completely separated and encoded with two-body stabilizers as in a repetition code. Logical qubits can appear in base qubit copies and in other parity qubits at the same time. An example to show this concept is shown in Fig.~\ref{fig:mixed_layout}. 

Note that some parts of the code, for example those with a repetition-like encoding, might require a higher qubit overhead in order to maintain the same code distance. This will reduce the overall space advantage of the parity code over a pure repetition code to a smaller factor. However, due to the flexibility gained from code deformation, this reduction can be kept to a minimum by always keeping the encoding as compact as possible during each step.

\section{Concatenated parity codes}\label{sec:concat-paritycode}
We now combine the bit-flip correcting properties of the parity codes with an additional encoding to protect against phase-flip errors. For this, we require the underlying qubits of the parity code to satisfy the following requirements:
\begin{enumerate}
    \item Phase-flip errors can be suppressed to a sufficient level without the bit-flip error rate growing too fast (such that phase-flips occur with a rate low enough to allow finite success probabilities of the full quantum algorithm, and bit-flips occur with a rate small enough to be correctable by the parity code).
    \item Initialization and readout in  $X$- and $Z$-bases, $Z$-flips, $R_x$ rotations,  CNOT and $R_{xx}$ or similar two-body gates with controllable angle, between nearest neighbors on a square-lattice, can be performed in a bias-preserving manner.
\end{enumerate}

An ideal example for such a system are cat qubits \cite{cai_bosonic_2021, terhal_towards_2020}, where the computational states can be defined via coherent states of an oscillator as
\begin{equation}
    |0_L\rangle = |\mathcal{C}_{\alpha}^+ \rangle =\frac{1}{\sqrt{2\left(1+e^{-2|\alpha|^2}\right)}}(|\alpha\rangle+|-\alpha\rangle)
\end{equation}

\begin{equation}
   |1_L\rangle  = \left|\mathcal{C}_\alpha^{-}\right\rangle=\frac{1}{\sqrt{2\left(1-e^{-2|\alpha|^2}\right)}}(|\alpha\rangle-|-\alpha\rangle).
\end{equation}
and thus
\begin{align}
    |+_L\rangle &\approx |\alpha \rangle \\
    |-_L\rangle &\approx |-\alpha \rangle.
\end{align}
The two coherent states with amplitude $\alpha$ and $-\alpha$  are not orthogonal, but their overlap exponentially decays with rate $\mathcal{O}( |\alpha|^2)$. This definition of a qubit in the oscillator Hilbert space is also called \textit{cat code}.

The subspace spanned by the above-defined cat qubit can be stabilized by employing well-known techniques realizing so-called \textit{driven-dissipative cat qubits} or \textit{Kerr cat qubits}. Driven-dissipative cat qubits are stabilized by engineering an oscillator with two-boson (e.g., two-photon or two-phonon) dissipation and a coherent two-boson drive \cite{mirrahimi2014dynamically}. Kerr cat qubits, instead, are stabilized by engineering a (non-linear) Kerr oscillator with a two-boson drive \cite{puri2017engineering}, which realizes a Hamiltonian spectrum with a gap between the subspace spanned by the cat qubit states and the excited states.

Such stabilized cat qubits exhibit a tunable noise bias: the phase-flip error is suppressed exponentially, while only a linear increase of the bit-flip error is observed~\cite{grimm_stabilization_2020,lescanne_exponential_2020} upon increasing the boson number. Furthermore, methods for constructing a universal set of logical gates using bias-preserving physical gates together with state preparation and measurement have been proposed for stabilized cat qubits~\cite{guillaud_repetition_2019, puri_bias-preserving_2020}. Consequently, stabilized cat qubits are a natural platform for biased-noise qubits.

To concatenate the cat qubits with a bit-flip correcting parity code in line with, e.g., Ref.~\cite{Fellner2022universal}, we defined the computational basis states such that bit-flip errors are dominant and phase-flip errors are suppressed. Note that in recent cat qubit literature (e.g.,  \cite{guillaud_repetition_2019, puri_bias-preserving_2020}) the computational basis states are often defined such that bit-flip instead of phase-flip errors are suppressed. Our basis choice is in alignment with previous literature on the parity encoding, and furthermore allows to represent the resulting logical gates (e.g. CZ) with commonly used terminology. However, one could also easily switch to a representation more natural to recent cat qubit terminology via a Hadamard transformation.

A concatenation of the phase-flip suppressing cat code and the bit-flip correcting parity code can protect against any type of error. For the concatenation, each base or parity qubit of the parity encoding is implemented as a cat qubit in a resonator. With this construction, bit-flip errors can be corrected using the $Z$-stabilizer operators of the parity code, while phase-flip errors are suppressed already at the level of each cat qubit.

When this code concatenation is used for fault-tolerant quantum computation, the available gate set is limited to (sequences of) physical operations which are bias-preserving in the cat code and transversal in the parity code.

Several bias-preserving physical operations are available for known cat qubit platforms~\cite{guillaud_repetition_2019, chamberland2022,puri_bias-preserving_2020}. Firstly, bias-preserving $X$ and $Z$ measurements and state preparations are available. Further, all single- or many-body rotations around the $X$-axis of the Bloch sphere, such as Pauli-$X$ rotations or $R_{xx}$ operations, trivially preserve bias against bit-flip noise. Additionally, a bias-preserving Pauli-$Z$ flip can be realized by adiabatically changing the phase of the two-boson drive from $0$ to $\pi$. Thereby, the quantum state undergoes a rotation through regions of the phase space of the oscillator which are outside of the qubit subspace. Similarly, CNOT gates and Toffoli gates can be implemented by changing the phase from $0$ to $\pi$ of the target cat qubit conditioned on the state of the control cat qubit(s). Refs.~\cite{guillaud_repetition_2019, chamberland2022,puri_bias-preserving_2020} introduce the code in a Hadamard-rotated basis, such that the $X$ and $Z$ gates introduced therein correspond to $Z$ and $X$ gates, respectively, in our case. The CNOT gate in Refs.~\cite{guillaud_repetition_2019, chamberland2022,puri_bias-preserving_2020} corresponds to a CNOT gate with control and target interchanged and the Toffoli gate is mapped to another unitary, $
{\rm {C}_{{\rm X}}{C}_{{\rm X}}Z=H^{\otimes3}\,CCX\,H^{\otimes3}}$, which applies a $Z$-flip whenever both control qubits are in the state $\ket{-}$, in our convention.

\section{Fault-tolerant gates}\label{sec:fault-tolerant-gates}

The most fundamental logical gates are the Pauli-$Z$ gate and the Pauli-$X$ gate which can be performed by applying $Z$ and $X$ operations as introduced in Section~\ref{sec:paritycode} to preserve the code space of the parity code, using the cat qubit's bias-preserving $Z$ and $X$ operations.

Some fault-tolerant logical gates can be implemented by a transversal application of single- or multi-qubit gates which are bias-preserving in the cat code, for example the CNOT gate. Other fault-tolerant logical gates can be constructed by magic state preparation and gate teleportation.

In the following, we will complement the set of natively fault-tolerant gates with a construction for fault-tolerant CNOT, CZ, H, S and HTH gates.

As main addition to this gate set, we will show how the fault-tolerant teleportation of S gates (or other $Z$ rotations) directly to parity qubits enables us to implement fault-tolerant many-body gates such as the CZ gate between arbitrary logical qubits locally, thus making use of the intrinsic logical connectivity of the parity code.

\subsection*{CNOT gate}

\begin{figure}
\begin{centering}
\includegraphics[scale=0.24]{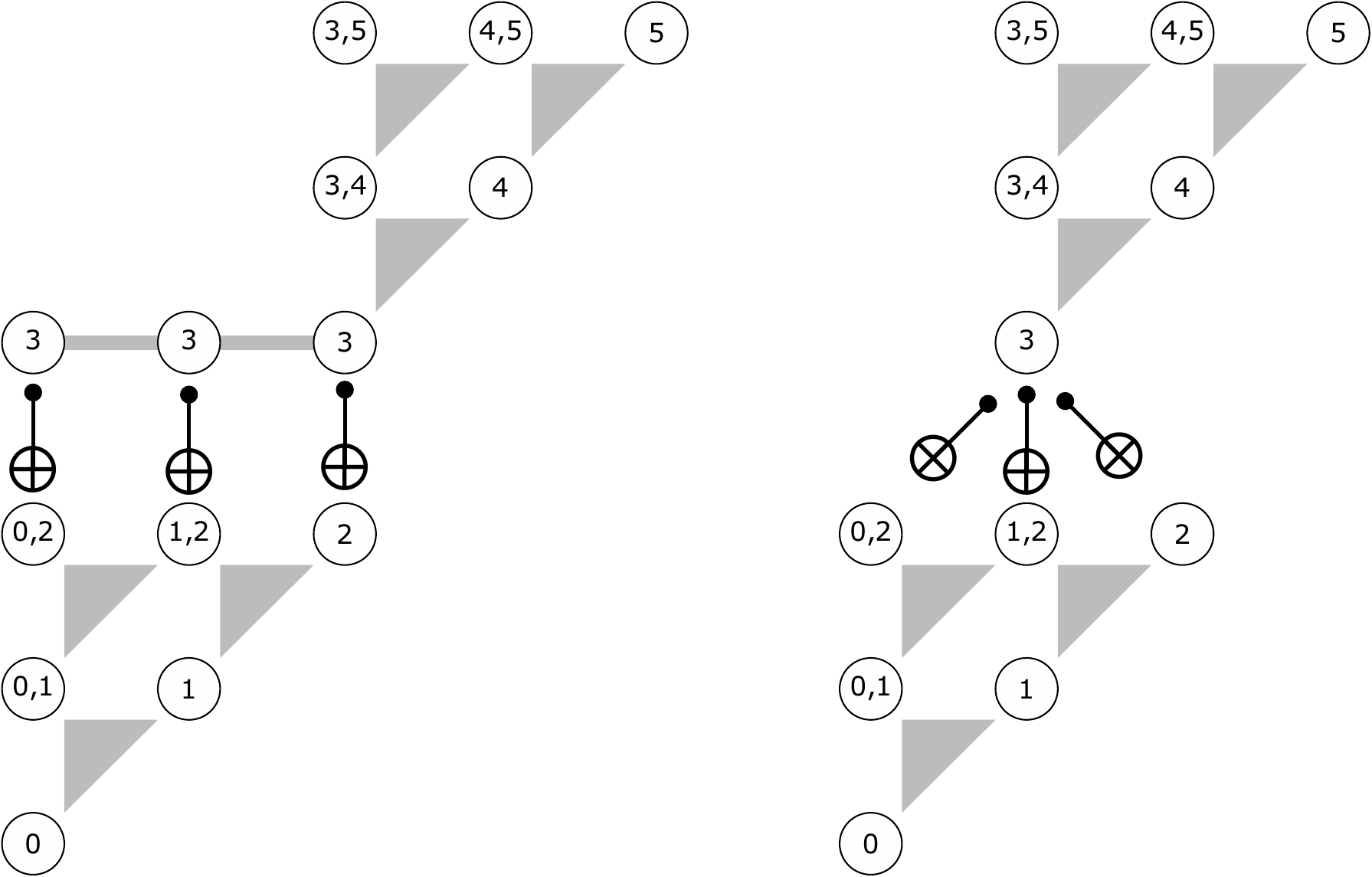}
\par\end{centering}
\caption{Schematic examples of a logical CNOT gate between logical qubits 3 (control) and 2 (target) from two different code blocks with all-to-all connectivity. All qubits with index 2 in their label must be targeted by exactly one CNOT gate and all control qubits must be labeled with the single index 3. Note, that the construction is still valid if additional qubits with  index 3 in their label exist. For a fault tolerant implementation (left), every CNOT must be controlled by a different qubit, which can be realized by adding copies of the same base qubit. The same control qubit could in principle be used for multiple CNOT gates (right), but this offers less fault-tolerance as a physical error of the control qubit propagates to multiple other qubits.\label{fig:Example-CNOT}}
\end{figure}

For a fault-tolerant CNOT gate one can use a transversal implementation as done for example in the repetition code~\cite{Nielsen2011}. As the target of a CNOT gate behaves like a Pauli-$X$ gate, the implementation must be such that every base and parity qubit with the logical target qubit's index must be targeted by one physical CNOT [see definition of logical Pauli-$X$, Eq.~\eqref{eq:logicalX}]. The action of the control part is in the Pauli-$Z$ basis, which means that every control must be on a qubit labelled exclusively by the logical control qubit index. The gate is still valid even if there are additional parity qubits which are not controlled by a physical CNOT, and in principle the different CNOT gates could also all be controlled by the same base qubit. However, in order to maintain transversality and avoid growing error chains, each CNOT should have its control on a separate copy of the corresponding base qubit\footnote{The fact that multiple qubits are controlled in this case does not come from a de-localization of the respective operators defining the action of the control side (logical $Z$ still corresponds to a single physical $Z$ on a base qubit), but from the de-localization of the logical $X$ operator on the target side.}. Figure~\ref{fig:Example-CNOT} shows examples of valid fault-tolerant and non-fault-tolerant CNOT constructions.

While such a transversal CNOT gate cannot always be applied between any pair of logical qubits in any code layout, a valid setup can always be obtained with appropriate code deformations. In particular, if there are too many parity qubits with the logical target's index, these can be removed from the code. If the corresponding parity qubits are inaccessible because they are surrounded by other qubits or too far away from the control qubits, this part of the code can be decoded and the qubits added at another part. Note that, in order to maintain a certain minimal code distance, it might be necessary to add new qubits before removing others. Finally, if not enough single-indexed control qubits are present, more can be added.

\begin{figure}
\begin{centering}
\includegraphics[scale=0.36]{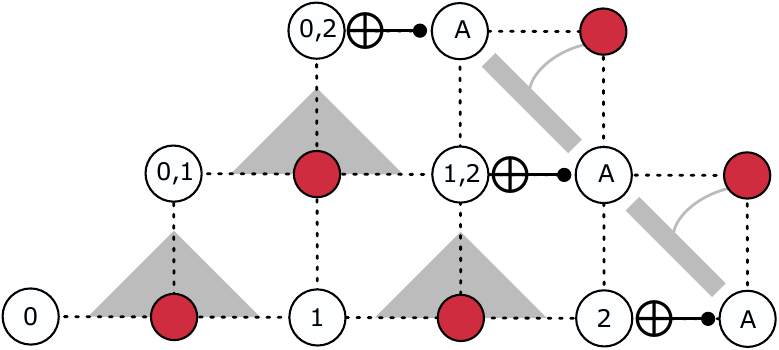}
\par\end{centering}
\caption{Transversal CNOT gate from an (ancilla) qubit A on shifted positions
to logical qubit 2. On a device with nearest-neighbor connectivity (dashed lines), usually only gates between measurement ancillas and parity qubits are natively available. To enable a local implementation of the CNOT gate, the parity qubits of logical qubit A can be placed on the positions usually occupied by measurement qubits and vice versa.\label{fig:local-CNOT}}

\end{figure}

If the underlying hardware is limited to nearest-neighbor connectivity on a square lattice, it might also be beneficial to position one of the two logical qubits with an offset such that parity qubits are on the positions of measurement ancillas and vice versa. This way, all physical CNOT gates are fully local and no SWAP gates or other decompositions are required. If, for example, the CNOT gate is used to inject a resource state from a logical ancilla qubit, the corresponding physical qubits which represent the logical ancilla can already be created on the offset position (see Fig.~\ref{fig:local-CNOT}).

\subsection*{S and T gates}

Universal quantum computing requires implementing rotations with arbitrarily small angles. This can be obtained for example with a decomposition using ${T=e^{i\pi/8\,Z}}$ gates. Similarly, the $S=e^{i\pi/4\,Z}=\sqrt{Z}$ gate has many useful applications, for instance for realizing the T gate via gate teleportation. Neither the S nor the T gate can be implemented directly in our code concatenation, but there exist gate teleportation protocols~\cite{gottesman1999,zhou2000} to implement them using a resource state which has been distilled to high fidelity using other error correction schemes~\cite{bravyi2005,steane1996,bravyi2012}. 
In particular, for implementations using biased-noise codes, Ref.~\cite{webster2015} describes a protocol to prepare high-fidelity ancilla states more efficiently than with traditional distillation schemes. The protocol uses physical controlled $Z$-rotations C$R_z$ on a platform with suppressed bit-flip noise to prepare the logical resource states ${\ket{+i} \propto \ket{0} + i\ket{1}}$ required for a logical S gate and ${\ket{T} \propto \ket{0} + \sqrt{i}\ket{1}}$ required for the T gate. In the proposed setup, all the required physical operations can be implemented in a bias-preserving manner and thus do not introduce any unexpected errors. In our setup (with suppressed phase-flip noise), this would translate to a preparation of $H\ket{+i}$ and $H\ket{T}$, with which the corresponding gates in our chosen basis, $HSH$ and $HTH$, can be performed using transversal logical CNOT gates, measurement and a conditioned correction.

Since the gates $HSH$ and $HTH$ comprise Pauli-$X$ components, an application of these gates to a logical qubit needs to involve all base qubits and parity qubits whose label contains the corresponding logical index. To avoid a complicated construction, the parity code may be deformed such that said logical qubit is represented by several copies of base qubits only.

Further, using ${H\ket{+i} = e^{-i\pi / 4}Z\ket{+i}}$, we note that we can also easily prepare the resource state required to perform the logical $S$ gate in our setup. 

\begin{figure*}
    \centering
    \includegraphics[scale=0.24]{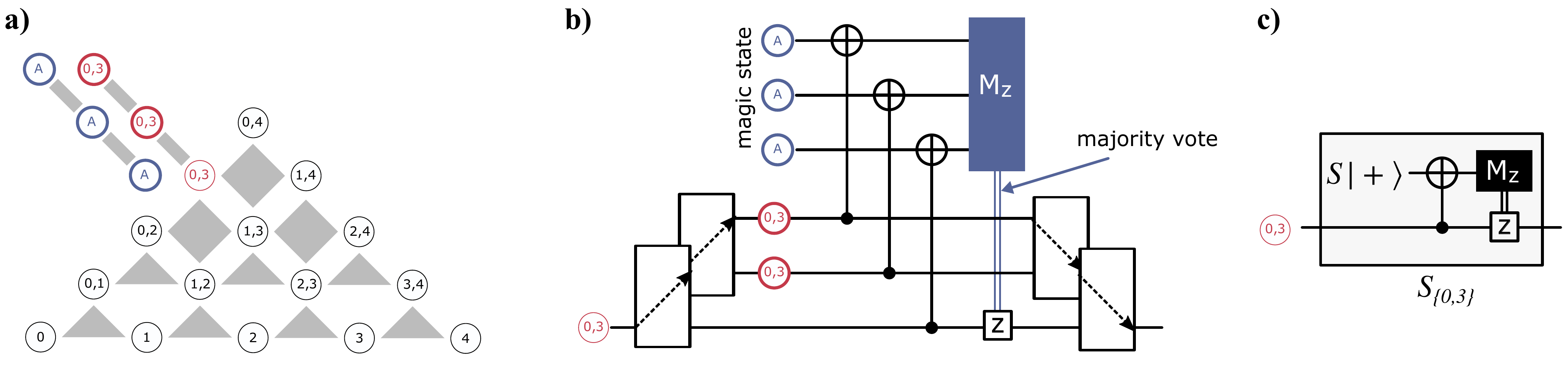}
    \caption{Example implementation of a logical $e^{i \pi/4 Z_0 Z_3}$ gate using a repetition encoding of a parity qubit and an ancilla in the state $S\ket{+}=\ket{+i}$. a) A possible layout for performing the gate. b) Circuit diagram for code distance ${d=3}$. If not already in the code, copies of the parity qubit ${0,3}$ are added to the code prior to the teleportation. The parity qubit can then be interpreted as encoded in a repetition code itself and the gate teleportation circuit can then be applied on the encoded parity qubit.
    After a transversal CNOT gate, the state of the logical ancilla qubit is measured fault-tolerantly by measuring all of its copies. This makes sure that any errors occurring during or after the CNOT can still be handled. After the teleportation, all but one of the copies of the parity qubit can be removed from the code again. Depending on the measurement outcome of the ancilla (the state that was measured in the majority of copies), a $Z$-flip operation is performed on the remaining parity qubit (this operation commutes with the removal of the other parity qubits, the $Z$-flip operation can be combined with any $Z$-flips resulting from the removal). Note that both the addition and the removal circuits for multiple parity qubits can be parallelized to a single round of measurements and corrections, respectively. c) The corresponding gate teleportation circuit without encoding and its effect on the parity qubit.}
    \label{fig:parity-S-gate}
\end{figure*}

\subsection*{CZ and $R_{zz}$ gates}

The parity encoding allows for a native implementation of the logical two-body rotation ${R_{zz}(\alpha)= e^{i \alpha/2 Z\otimes Z}}$  (and similarly also higher-order many-body rotations) using a single-body rotation on the corresponding parity qubit~\cite{Fellner2022universal}. As arbitrary small-angle $Z$ rotations are not bias-preserving, this construction cannot be directly used for a fault-tolerant implementation. However, with distillation and injection~\cite{webster2015}, it is possible to implement a fault-tolerant S gate on a parity qubit and, in further consequence, a logical ${R_{zz}(\pi/2)= e^{i \pi/4 Z\otimes Z}}$ rotation (see Fig.~\ref{fig:parity-S-gate}). Fig.~\ref{fig:parity-S-gate}c) shows the gate teleportation protocol used to perform an S gate on a parity qubit using a resource state. To perform this operation fault-tolerantly with repetition-encoded magic state ancillas, one must first create enough copies of the parity qubit of interest such that the parity qubit itself can be understood to be encoded in a repetition code as well. Then, we can use the teleportation scheme with a transversal CNOT gate between all the copies of the parity qubit and the copies of the magic state ancilla. The redundant copies of the parity qubit can be removed again after the process. In fact, this can even be done already during the measurement of the teleportation protocol, by also measuring all the other redundant qubits at the same time.

Analogously to Ref.~\cite{Fellner2022universal}, a logical CZ gate acting on two logical qubits $i$ and $j$ can be performed by applying the above-introduced S-gate teleportation scheme onto a parity qubit with indices $i$ and $j$, and applying an S-gate teleportation scheme to the base qubits $i$ and $j$. Using other magic states, for example $|T\rangle$, also smaller many-body rotations can in principle be implemented following the same idea. 

The implementation of such gates directly on parity qubits has the advantage that, as long as enough magic ancillas are available, many long-range interactions can be implemented in parallel without the need for qubit routing: The qubit overhead for implementing such interactions originates solely from the preparation of magic states. In principle, there is no limitation on the amount of parallel interactions or the non-locality of the interactions, as long as there is a parity qubit for each interaction and enough space to create magic state ancillas. 
In some cases, additional qubit overhead is necessary in order to place all the required qubits with local stabilizers only. For many problem settings, this overhead is rather low  \cite{terhoeven2023} and outweighs the overhead of alternative approaches, for example via qubit routing.
For example, the LHZ layout is still an efficient layout, even when some of the parity qubits are not required for interactions.
However, instead of trying to use the parity encoding for everything, we stress that it is important to assess on a case-by-case basis whether or which parts of an algorithm should be implemented on parity qubits, and in which cases other approaches are advantageous.

\subsection*{Hadamard gate}

A logical Hadamard gate has been proposed for bit-flip protected cat qubits and phase-flip correcting repetition code setup using teleportation of a specific resource state with a Toffoli gate, see Ref.~\cite{guillaud2021}. One can use the same scheme in the basis used in our work, as the resulting transformed gate is still a Hadamard, ${HHH = H}$.

Alternatively, a logical Hadamard gate can be realized also with a single logical ancilla qubit by implementing consecutive gate teleportation schemes for the gates $S^\dagger$ and $SHS$ using the relation ${H=S^\dagger SHS S^\dagger}$, as proposed in Ref.~\cite{chamberland2022}.

Together with the logical CZ gate, the Hadamard gate can be used to replace the direct, transversal implementation of the CNOT gate and thus avoid any locality restrictions.

\section{Conclusion}\label{sec:conclusion}

The parity code can be understood as a particular variant of LDPC code which extends beyond pure error-correcting capabilities, bringing in powerful and flexible tools for logical gates and compact encodings.
Together with biased-noise qubits as for example cat qubits, the presented code concatenation is a promising candidate for early realizations of fully fault-tolerant quantum algorithms of practical use.

We demonstrated how to implement a logical CZ gate acting on two logical qubits while only locally affecting base or parity qubits corresponding to the logical qubits. Due to this unique feature of our parity code, logical CZ gates, but also more general $R_{zz(...)}$ rotations are easily parallelizable without any limitations from routing. The parity code thus has the potential for the efficient implementation of many different quantum algorithms containing such gates.

Further, we showed how to implement a universal gate by combining operations applied directly to parity qubits with transversal gates and gate teleportation methods. The parity code can be flexibly transformed by code deformation in order to facilitate or enable the implementation of a sequence of these logical gates to perform arbitrary quantum computations.

Advantageously, our proposals for the logical gate implementations rely on a deformation of the parity code itself and can therefore be implemented without any additional requirements from the hardware setup. Only nearest-neighbor interactions on a 2D grid and measurements are required for the deformation operations.

Compared to the repetition code, the parity code has a higher encoding rate, the qubit overhead compared to the repetition code is reduced by a factor of approximately 2 in the limit of large numbers of logical qubits. At the same time, the parity code has relatively simple stabilizers which can be easily measured on a device with square-lattice connectivity.

Comparable classical codes such as the optimized LDPC codes from Ref.~\cite{ruiz2024} exhibit a higher encoding rate than the parity code at the cost of more complicated stabilizers which involve next nearest neighbor interactions and overlapping qubit connections in the physical implementation. Due to the more complicated stabilizers, however, transversal gates and code deformation are technically more difficult to achieve with these codes, and hence a single-layer implementation is much less efficient, if not technically intractable.

In the future, a more thorough comparison of the error-correcting capabilities is needed, since good error-correction depends on many aspects beyond the code distance and the encoding rate. For instance, decoding strategies for syndrome measurements can be investigated in more detail in future.

Finally, our approach can be extended to combining the parity code with other biased-noise codes instead of using cat qubits for the noise-bias.

\section{Acknowledgements}
  The authors thank Christophe Goeller for helpful discussions and comments on the manuscript.
  This study was supported by the Austrian Research Promotion Agency (FFG Project No. FO999909249, FFG Basisprogramm).
  This research was funded in part by the Austrian Science Fund (FWF) SFB BeyondC Project No. F7108-N38, START grant under Project No. Y1067-N27 and I 6011. For the purpose of open access, the author has applied a CC BY public copyright licence to any Author Accepted Manuscript version arising from this submission. This project was funded within the QuantERA II Programme that has received funding from the European Union's Horizon 2020 research and innovation programme under Grant Agreement No. 101017733.

\bibliographystyle{apsrev4-2}
 \input{compiled_bib.bbl}

\end{document}

%% file: compiled_bib.bbl
%